\documentclass[sigconf]{acmart}

\usepackage{tikz}
\usepackage{amsmath}

\usepackage{algorithm}
\usepackage{algpseudocode}
\usepackage{listings}
\definecolor{backcolour}{rgb}{0.95,0.95,0.92}
\definecolor{codegray}{rgb}{0.5,0.5,0.5}
\lstdefinestyle{mystyle}{
    backgroundcolor=\color{backcolour},   
    numberstyle=\tiny\color{codegray},
    basicstyle=\ttfamily\footnotesize,
    breakatwhitespace=false,         
    breaklines=true,                 
    captionpos=b,                    
    keepspaces=true,                 
    numbers=none,
    numbers=left,
    numbersep=5pt,                  
    showspaces=false,                
    showstringspaces=false,
    showtabs=false,                  
    tabsize=2
}
\algdef{SE}
[CLASS]
{Class}
{EndClass}
[1]
{\textbf{class} \textsc{#1}}
{\textbf{end class}}
\usepackage{makecell}
\usepackage{subcaption}
\usepackage[framemethod=tikz]{mdframed}
\newmdenv[
outerlinewidth = 1,
roundcorner = 10pt,
backgroundcolor = gray!10,
outerlinecolor = blue!20,
innertopmargin = \topskip,
splittopskip = \topskip
]{obsBox}

\newcommand\FramedBox[3]{%
  \setlength\fboxsep{5pt}
  \fbox{\parbox[t][#1][c]{#2}{\small #3}}}

\def\BibTeX{{\rm B\kern-.05em{\sc i\kern-.025em b}\kern-.08em
    T\kern-.1667em\lower.7ex\hbox{E}\kern-.125emX}}

\usepackage{multirow}
\newcommand{\bacfuzz}{\textbf{BACFuzz}}

\usepackage{enumitem}

\copyrightyear{2025}
\acmYear{2025}
\acmDOI{XXXXXXX.XXXXXXX}
\acmConference[Under peer-review]{XXX}{June xx--xx,
  2025}{US}

\begin{document}


\date{}

\title{\bacfuzz: Exposing the Silence on Broken Access Control Vulnerabilities in Web Applications}




\author{I Putu Arya Dharmaadi}
\affiliation{%
  \institution{University of Groningen}
  \city{Groningen}
   \country{Netherlands}
}
\email{arya.dharmaadi@rug.nl}

\author{Mohannad Alhanahnah}
\affiliation{%
  \institution{Chalmers University}
  \city{Gothenburg}
  \country{Sweden}
}
\email{mohannad.alhanahnah@chalmers.se}

\author{Van-Thuan Pham}
\affiliation{%
  \institution{The University of Melbourne}
  \city{Melbourne}
  \country{Australia}
}
\email{thuan.pham@unimelb.edu.au}

\author{Fadi Mohsen}
\affiliation{%
  \institution{University of Groningen}
  \city{Groningen}
\country{Netherlands}
}
\email{f.f.m.mohsen@rug.nl}


\author{Fatih Turkmen}
\affiliation{%
  \institution{University of Groningen}
  \city{Groningen}
  \country{Netherlands}
}
\email{f.turkmen@rug.nl}

\renewcommand{\shortauthors}{Dharmaadi et al.}

\begin{abstract}

Broken Access Control (BAC) remains one of the most critical and widespread vulnerabilities in web applications, allowing attackers to access unauthorized resources or perform privileged actions. Despite its severity, BAC is underexplored in automated testing due to key challenges: the lack of reliable oracles and the difficulty of generating semantically valid attack requests. We introduce \bacfuzz, the first gray-box fuzzing framework specifically designed to uncover BAC vulnerabilities, including Broken Object-Level Authorization (BOLA) and Broken Function-Level Authorization (BFLA) in PHP-based web applications. \bacfuzz~combines LLM-guided parameter selection with runtime feedback and SQL-based oracle checking to detect silent authorization flaws. It employs lightweight instrumentation to capture runtime information that guides test generation, and analyzes backend SQL queries to verify whether unauthorized inputs flow into protected operations. Evaluated on 20 real-world web applications, including 15 CVE cases and 2 known benchmarks, \bacfuzz~detects 16 of 17 known issues and uncovers 26 previously unknown BAC vulnerabilities with low false positive rates. All identified issues have been responsibly disclosed, and artifacts will be publicly released.
\end{abstract}



\maketitle

\section{Introduction}
The widespread adoption of web applications has brought associated security risks to the forefront, drawing significant attention from both academia and industry. To address these risks, OWASP has published two influential guides: the OWASP Top 10 Web Application Security Risks \cite{OWASP_Top10_2024} and the OWASP API Security Top 10 \cite{noauthor_owaspAPITOP10_nodate}. Last updated in 2023, both documents identify Broken Access Control (BAC) as the most prevalent and critical security flaw.

BAC enables attackers to access unauthorized functionality or data—often resulting in vertical privilege escalation \cite{PortSwigger_Accesscontrol_2024}. For example, a regular user might overwrite an administrator’s email address or invoke privileged actions such as deleting user accounts. Several recent high-impact incidents involving BAC vulnerabilities (see Section~\ref{real-cases}) underscore the urgency of robust access control testing.

While OWASP provides guidelines for testing access control \cite{OWASP_WASTG_2024}, automated tools capable of doing so with minimal human effort remain lacking. Manual analysis is labor-intensive and difficult to scale for complex applications \cite{liang_fuzzing_2018}, and most existing fuzzers primarily target crash-inducing bugs \cite{zhu_fuzzing_2022, beaman_fuzzing_2022}. Recent surveys \cite{golmohammadi_testing_2023, zhang_open_2023, dharmaadi_fuzzing_2025} confirm that fuzzing research rarely addresses logical vulnerabilities such as BAC.

Our analysis of BAC characteristics (see Section~\ref{preliminary-analysis}) identifies two main challenges that hinder fuzzing for BAC. First, designing a test oracle—i.e., determining whether a request triggers a vulnerability—is difficult because BAC violations typically do not cause crashes or produce explicit error messages. Detecting such silent flaws often requires deep domain knowledge, as seen in recent research on SQL injection \cite{trickel_toss_2023}, excessive data exposure \cite{pan_edefuzz_2024}, and DNS resolver bugs \cite{zhang_resolverfuzz_2024}. Heuristics based on HTTP response codes or messages are often unreliable due to ambiguous or generic server feedback. Second, API requests often include many parameters with diverse data types and dependencies, resulting in a vast search space. Generating inputs that are both valid and semantically meaningful is non-trivial, as malformed requests are usually rejected.

To address the oracle challenge, we draw inspiration from prior work on injection flaws \cite{trickel_toss_2023}. Once a request is accepted, the database-backed web application typically parses it, performs validation and authorization checks, and then constructs backend SQL queries. If a mutated input value appears in a resulting query, it suggests that some validation—potentially an authorization check—was bypassed. While this is a necessary condition for BAC, sufficiency depends on the context; for example, in Broken Function Level Authorization (BFLA) cases, the operation may not even be visible in the current user's interface. Based on this insight, we define a set of rules to reliably detect BAC violations (see Section~\ref{SQL-checking}).

To address the input generation challenge, we focus on the two most common BAC types: Broken Object Level Authorization (BOLA) and Broken Function Level Authorization (BFLA). Rather than mutating all parameters equally, we target those referencing protected objects or functionalities—i.e., those likely to reveal privilege differences between users. We collect traffic from users with different roles, label the requests, and leverage Large Language Models (LLMs) to identify semantically important parameters. LLMs are well-suited to this task due to their reasoning capabilities and ability to understand structured data. This significantly reduces the mutation space while preserving request validity.

Building on these ideas, we present \bacfuzz, the first automated fuzzing tool specifically designed to detect BAC vulnerabilities in PHP-based web applications. \bacfuzz~uses a grey-box fuzzing approach, leveraging runtime feedback collected via lightweight code instrumentation—a method widely regarded as effective and scalable in modern fuzzing \cite{manes2019art}. Our instrumentation uses function hooking techniques to monitor original PHP functions related to SQL queries, enabling precise feedback and oracle validation.

We evaluate \bacfuzz~on 20 PHP-based Web Under Test (WUT) applications. It successfully reproduces 16 of 17 known BAC issues and uncovers 26 previously unknown vulnerabilities, all of which have been responsibly disclosed. While instrumentation introduces minimal overhead, it enables the generation of semantically valid, high-coverage test cases—frequently bypassing 4xx response rejections. Our results demonstrate that \bacfuzz~is effective, scalable, and practical for uncovering BAC vulnerabilities in real-world applications.

\noindent\textbf{In summary, this paper makes the following contributions:}
\begin{enumerate}[leftmargin=*]
\item \textbf{Novel Approach}: We propose the first grey-box fuzzing method specifically designed to automatically detect object- and function-level BAC vulnerabilities in web applications.
\item \textbf{New Exploitation Strategy}: We introduce an active checker module that applies advanced grey-box fuzzing techniques, including LLM-guided parameter analysis and reference mutation, to exploit captured HTTP requests.
\item \textbf{Novel Oracle Verification}: We develop a verification module that inspects SQL queries issued by the application. If mutated inputs appear in these queries and vulnerability type-specific conditions are met, we confirm that the target object or function is vulnerable.
\item \textbf{New BAC Vulnerability Dataset}: We curate and release a dataset of web applications with known CVE-reported BAC vulnerabilities, addressing the lack of existing benchmarks for BAC detection tools.
\end{enumerate}

\section{Background and Motivation}
\label{motivation}

\subsection{Background}
\label{background}
According to the definition provided by OWASP  \cite{OWASP_Top10_2024}, BAC refers to a vulnerability that occurs when a web application allows a certain access request to read or modify resources that it should not be able to. One specific example of BAC is \textbf{IDOR (Insecure Direct Object References)}, which enables malicious parties to access protected resources by sending resource identifiers through user-controlled parameters. 
There are two risks related to IDOR, as follows.



\paragraph{BOLA---Broken Object Level Authorization (API1:2023)}
BOLA occurs as inadequate access controls allow a user to manipulate or access data objects that they are not permitted to view or modify. For example, assume that a web portal application stores \textit{products} from different sellers and only the corresponding seller is allowed to modify them. When a seller can modify a product that belongs to another seller by sending the ID reference of the product, the web application is said to have BOLA. Another example is explained in Figure \ref{fig:cve-example}.

\paragraph{BFLA---Broken Function Level Authorization (API5:2023)}
BFLA occurs when a web application fails to enforce proper authorization checks on user access to specific web functions (e.g., API endpoint). For example, consider the case in which a \textit{button to create a new object} on a web page is removed when non-admin users access the page since only administrators are allowed to create a new object. The application is said to have BFLA if the web server accepts/executes a direct HTTP request originating from non-admin users to create an object.

\subsubsection*{BOLA vs BFLA} When certain HTTP requests are supposed to be available only for some users, and the server executes the same requests even if they originate from other users, it is \textbf{BFLA}. On the other hand, \textbf{BOLA} arises when the request is indeed available for a user yet the user modifies a parameter identifying an object with another ID reference that is not available his/her pages, and the server executes the request.


\begin{figure}[!t]
    \FramedBox{2.1cm}{0.45\textwidth}{\textbf{POST} /wp-admin/admin-ajax.php HTTP/1.1
    \\ \textbf{Host}: example.com
    \\ \textbf{Content-Type}: application/x-www-form-urlencoded
    \\
    \\ action=wcfm\_ajax\_controller\&controller=wcfm-customers-manage\&..............................................\textbf{customer\_id}=\textcolor{red}{\textbf{1}}......................................
    }
    \caption{An HTTP Request exploiting CVE-2024-8290. Normally, the script on the web client sets the \textit{customer\_id} parameter to have a default value of 0, meaning the creation of a new customer. Arbitrarily changing the value to 1 leads the submitted data to replace the original data of a user with the ID of 1.}
    \Description{An HTTP Request exploiting CVE-2024-8290. Normally, the script on the web client sets the \textit{customer\_id} parameter to have a default value of 0, meaning a new customer creation. Arbitrarily changing the value to 1 leads the submitted data to replace the original data of a user with the ID of 1.}
    \label{fig:cve-example}
    \vspace{-0.3cm}
\end{figure}

\subsection{Motivating Examples}
\label{real-cases}
This section presents two real-world examples of BAC-related vulnerabilities to motivate the relevance of automated BAC vulnerability detection. 

CVE-2024-8290 \cite{cve_8290_2025}, disclosed in September 2024, represents a critical BAC vulnerability within the \textit{wc-frontend-manager} plugin for WordPress. 
This CVE report has a significant impact on the score (CVSS 8.8 with high severity), affecting approximately 20,000 users, according to the report from WordFence \cite{wordfence_wp-affected_2024}.
This vulnerability arises from inadequate validation of the ID parameter in the 
plugin function that stores a new or a modified customer object. Although the function verifies user capabilities, making only certain roles allowed to call the function, it fails to restrict the scope of objects that can be modified. As a result, an authenticated attacker with lower privileges, such as managers, can exploit this flaw to alter the email addresses of administrator accounts by providing valid administrator IDs (see Figure \ref{fig:cve-example}). After recording her email address, the attacker can call password resets, gain unauthorized access to administrative accounts, and potentially take full control of the affected WordPress site.

Another example is CVE-2024-7437 \cite{cve_2024_7437}, which happens in the SMF (Simple Machines Forum) application. In this application, BAC occurs when a malicious user alters the vulnerable parameter (i.e., \textit{alert ID}) in the request, allowing him to delete other users' alerts. This vulnerability is similar to the one in WordPress, as improper authorization checks on client-submitted object references cause both.

\begin{table*}[htbp]
    \centering
    \caption{Collection of CVEs reporting BAC, sorted by affected App size (the number of LoC). For simplicity, HTTP Param refers to name-value pairs placed in the HTTP header, the URL query string, or the HTTP body.}
    \label{tab:existing_CVEs}
    \vspace{-0.3cm}
    \begin{tabular}{c|c|c|c|c}
    \hline
        CVE No. & Affected App (+ Plugin) & Method & \makecell{Vulnerable URL} & \makecell{Vulnerable Param}\\
    \hline
        CVE-2025-0843 & Library Card System & GET & /del.php & del=<<student\_id>> \\
        CVE-2025-3536 & Employee Management System & GET & /admin/delete-user.php & id=<<user\_id>> \\
        CVE-2025-3537 & Employee Management System & POST & /admin/update-user.php & user\_id=<<user\_id>> \\
        CVE-2024-55231 & Notes Sharing Management System & POST & /user/edit-notes.php & editid=<<id>> \\
        CVE-2024-55232 & Notes Sharing Management System & GET & /user/manage-notes.php & delid=<<id>> \\
        CVE-2024-40480 & Online Exam System & GET & /admin/update.php & uemail=<<email>> \\
        CVE-2025-0802 & Best Employee Management System & POST & /admin/Operation/User.php & del\_id=<<user\_id>> \\
        CVE-2023-46449 & Inventory Management System & POST & ../action/edit\_update.php & user\_id=<<user\_id>> \\
        CVE-2024-3139 & Computer Lab Management System & POST & /classes/Users.php & f=save id=<<id>> \\
        CVE-2024-9082 & Online Eyewear Shop & POST & /classes/Users.php & f=save id=<<id>> \\
        CVE-2024-7658 & Projectsend  & GET & /process.php & do=get\_preview\&file\_id=<<id>> \\
        CVE-2024-7437 & Simple Machines Forum & GET & /index.php & do=remove;aid=<<alert\_id>> \\
        CVE-2024-7438 & Simple Machines Forum & GET & /index.php & do=read;aid=<<alert\_id>> \\
        CVE-2024-8290 & WordPress + WC + WCFM & POST & /wp-admin/admin-ajax.php & customer\_id=<<admin\_id>>\\        
        CVE-2023-43663 & Prestashop & GET & .../action/disable/<<module\_id>> & \\
    \hline
    \end{tabular}
\end{table*}

\section{Preliminary Analysis of BAC Characteristics}
\label{preliminary-analysis}
To gain deeper insights into the characteristics and distribution of BAC-related vulnerabilities in real-world software systems, we performed a preliminary investigation of the BAC attributes associated with publicly disclosed CVEs from recent years.

\subsection{Data Collection}
Firstly, as explained in Section~\ref{background}, we use \textit{Broken Access Control} and \textit{IDOR} as the primary keywords in our search to query the CVE repository \cite{CVE} for reports published in the last three years (between April 2022 and April 2025). 
We further restricted our CVE selection to cases that occurred in open-source web applications. This constraint ensures that we can reproduce the vulnerable behaviour in a local environment and inspect the source code if necessary. Projects that required commercial licenses, closed-source systems, or non-web contexts were excluded. In addition, since the grey-box web fuzzer we design requires platform-specific instrumentation, we limit our focus to CVEs affecting PHP-based applications, as PHP remains one of the most widely used platforms. Finally, to ensure the presence of BAC vulnerability, we manually reviewed each CVE entry and included only those with sufficient technical detail (e.g., vulnerable URL and parameters) to reproduce the issue.

\subsection{Challenges to Reveal BAC}
\label{root-cause-analysis}
We collected 15 CVEs (see Table~\ref{tab:existing_CVEs}) after applying the methodology described in the previous section. We then manually analyzed the vulnerable functions and objects to find ways of triggering these vulnerabilities through the web UI with the aim of identifying relevant HTTP requests to submit and verifying successful trigger/exploitation. Since our goal is to design a fuzzer tailored to BAC vulnerabilities, we identified the challenges for automation of the process.


\subsubsection*{C1: Identifying Potentially Vulnerable Functions}
\label{identify-vulnerable-function}
Web applications dynamically generate function calls based on user context, session state, or permissions, making the discovery process heavily context-dependent. Furthermore, they may rely on aliasing mechanisms in their APIs, making the server file names irrelevant to available endpoints. Consequently, traditional methods, such as scanning for files on the server and invoking them through HTTP requests can be less effective. 

Consider CVE-2023-43663 (see Table \ref{tab:existing_CVEs}), which affects the Presta-Shop application (has 8.6k stars on Github) \cite{prestashop}, as an example of this observation. The vulnerable function of the Prestashop application lies in the \textit{AdminDashboardController.php} file, which is not visible in the URL. To trigger that function, rather than calling the file name via URL, a non-admin user should call a certain URL ending with \textit{/disable/«module\_id»}. Failing to produce the correct URL with the correct \textit{module\_id} prevents the fuzzer from reaching the vulnerable function.

In addition to collecting all functions, flagging some of the functions as potentially vulnerable is also challenging. Popular security tools, such as OWASP ZAP \cite{noauthor_owaspzap_nodate} and Burp Suite \cite{noauthor_burp_nodate}, solve this challenge by requiring users to manually configure an access control list (ACL) that defines
which functions should be allowed or denied for specific users. However, such reliance on human intervention undermines the applicability of that solution for fuzzing, which aims to minimize manual effort.

\subsubsection*{C2: Identifying Potentially Vulnerable Objects}
\label{allowed-protected-object}
Differentiation of allowed and restricted objects is crucial to precisely attempt to 
trigger BOLA vulnerabilities. 
Due to the context-dependent nature of the objects, opening a webpage with a specific user role and observing visible objects provides a baseline for identifying resources accessible to that role. However, the real challenge lies in the discovery of restricted objects, particularly those with differentiated access permissions (e.g., accessible for READ operations but restricted for UPDATE or DELETE actions). In addition, the restricted objects are frequently hidden behind indirect references or system-generated values that are only visible in the body of the HTTP request and therefore are difficult to observe without manual inspection.

For example, in CVE-2024-8290 and CVE-2024-7437, restricted objects are related to the administrator and alert, and are referred using \textit{admin ID} and \textit{alert ID}, respectively. Failing to identify the parameter names and generate the correct IDs prevent the web fuzzers from triggering the vulnerability. 

\subsubsection*{C3: Oracle Verification}
\label{oracle-verification}
Semantic bugs like BAC do not easily manifest themselves in observable states (e.g., crash, memory corruption, etc). Therefore, confirming the occurrence of the BAC vulnerability is hard because there is no clear signal from WUT to confirm, especially in the case of unexpected resource modification. Solely relying on web response messages can be less effective since the WUTs may reply with unclear information. 

For example, in CVE-2025-3536, when a malicious user calls the vulnerable URL with a correct \textit{user\_id}, the server replies with redirected responses without a clear success message. Another example, in CVE-2024-8290, when calling the vulnerable URL with an unexisting \textit{admin\_id}, a user receives a clear success message even though the server does not execute the update admin request due to the incorrect \textit{id}.
Therefore, the fuzzer needs accurate information that cannot be obtained from the response messages.

\subsubsection*{C4: Reducing Randomness}
\label{exploitation-strategies}
In general, web applications expose a large number of endpoints with numerous parameters in the request headers, URL and body. This situation makes random selection of one of these inputs and its alteration with a byte array or random strings less effective because there are huge input spaces to explore. Normally, most web servers apply certain input validation rules, making them often reject abnormal inputs. To make fuzzing more effective, certain endpoints and parameters that have a greater chance of triggering the access control vulnerability should be prioritized. For example, in CVE-2024-8290, mutating the value of the \textit{name} parameter is unlikely to trigger a BAC vulnerability, whereas mutating the \textit{customer\_id} parameter does.

In addition, web fuzzers should reduce the use of completely random values in order to increase chances of reaching and triggering vulnerable code. 
For example, in CVE-2024-40480, filling the \textit{uemail} with a random value or a grammatically valid value that does not exist in the database will not trigger BAC. Therefore, to work efficiently in revealing BAC, a fuzzer should use fewer random values.

\subsection{Scope of the Work}
\label{scope-of-work}
Based on the insights gained from the preliminary analysis, we define the scope and assumptions that underlie the proposed approach.

\begin{enumerate}[leftmargin=*]
    \item Given the wide range of BAC cases, our work focuses only on role-based access control (RBAC) \cite{zelkowitz_role-based_1998}, arguably one of the most commonly implemented access control models, which allows or prohibits certain users from accessing or modifying any function or object based on their roles. Other models, such as context-based or attribute-based access control \cite{10.1145/3007204}, which regulate whether users are allowed to access certain objects based on the state of the users, are out of the scope.
    
    \item Our work is limited to uncovering BAC at the code level. 
    Issues caused by design-time errors 
    (e.g., a user with limited privileges is assigned a role beyond her privileges) are out of the scope. We assume that such errors do not reflect a vulnerability in the code but in the instantiation of the RBAC model (i.e., user-role and role-permission assignments).

    \item The proposed fuzzer does not explore available actions to insert data in the beginning. We assume the human tester normally prepares the WUT with some initial data, including registered users with different roles, before performing security testing.
\end{enumerate}

\section{Proposed Approach}
\label{proposed-approach}
Based on the challenges described in Section \ref{root-cause-analysis}, we propose three techniques tailored to BAC vulnerability detection: hierarchical role analysis, reference mutation, and SQL checking. These techniques are then implemented in a fuzzer we call \bacfuzz ~(Section \ref{sec:implementation}) in revealing BAC.



\subsection{Hierarchical Role Analysis}
\label{request-interception}
For vulnerable function identification (C1), we propose hierarchical role analysis to find HTTP requests that only appear in higher-role users. Firstly, this process makes the fuzzer open web pages with a specific user role, navigate the pages, fill forms, click buttons and links, and save HTTP requests the browser sends. The fuzzer repeats these actions for all available user roles. All saved requests replied to with a valid response code (i.e., 2xx) are stored in the \textbf{request corpus} grouped by their roles. Furthermore, param-value pairs are extracted from URL queries and body payloads and stored in another corpus (i.e., \textbf{param corpus}). To determine whether a request is new and unique, the fuzzer compares the request URL, request method, URL queries (only param name without value), and body payloads (only param name without value).

The request that appears in certain user roles yet disappears in the other roles is marked as a potentially vulnerable function. Then, the fuzzer can focus on \textbf{BFLA testing} (also called \textit{vertical access control testing}) on these request types by sending the requests using a lower-role account. As explained in Section \ref{identify-vulnerable-function}, the vulnerable function in CVE-2023-43663 can be triggered by calling the link using a lower-role account. For the other requests that are not marked, the fuzzer conducts \textbf{BOLA testing} (also called \textit{horizontal access control testing}) by applying reference mutation (Section \ref{reference-mutation}).

\subsubsection*{\textbf{Role Labelling}.}
The fuzzer labels each request with role names that the fuzzer uses to trigger the request. For example, the fuzzer logging in with an admin account can execute some actions on the admin page. The requests collected during this session are labelled with admin. When the same request comes from another fuzzer instance logging in with a different role (e.g., manager), the fuzzer only adds a new label (i.e., manager) to the request in the corpus. In addition, the fuzzer does the same labelling process for param-value pairs stored in the param corpus.


\subsection{Reference Mutation}
\label{reference-mutation}
To trigger BOLA vulnerabilities from collected requests, we propose a reference mutation. This proposed method involves sys-gen data collection, reference parameter analysis, and value alteration.

\subsubsection{Sys-Gen Data}
\begin{figure}[t]
     \begin{subfigure}[b]{0.5\textwidth}
         \centerline{\includegraphics[width=0.9\linewidth]{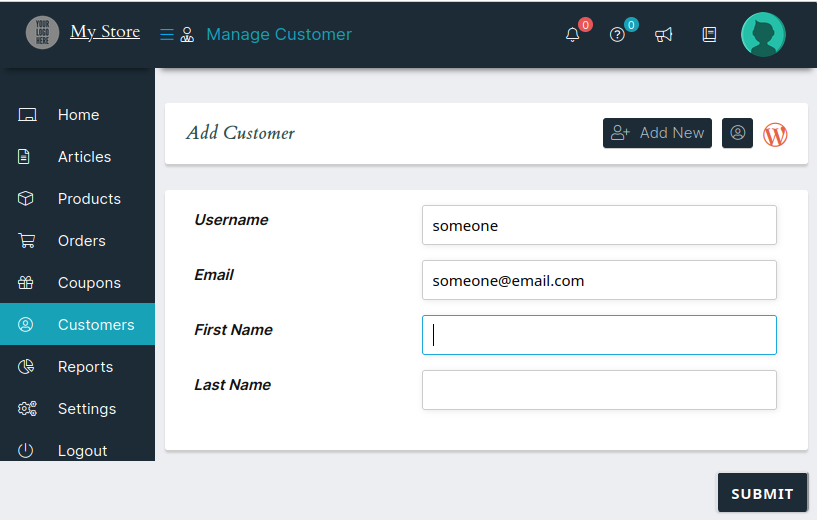}}
         \caption{Add customer form}
         \label{fig:form}
     \end{subfigure}
     \hfill
     \\
     \begin{subfigure}[b]{0.5\textwidth}
         \centerline{\includegraphics[width=0.8\linewidth]{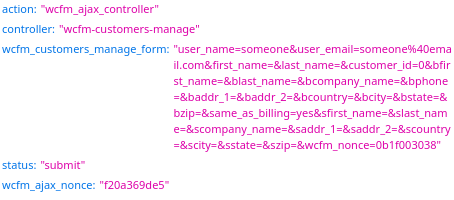}}
         \caption{HTTP Request Body}
         \label{fig:http-request}
     \end{subfigure}
    \caption{In CVE-2024-8290, after a user fills in a form (\ref{fig:form}) and clicks the submit button, the browser generates an HTTP request with body payload (\ref{fig:http-request}) containing additional fields called sys-gen data: \textit{action}, \textit{controller}, \textit{status}, and \textit{wcfm\_ajax\_nonce}, which is invisible in users' direct view. Furthermore, the field of \textit{wcfm\_customers\_manage\_form} also consists of some sys-gen data. 
    }
    \Description{In CVE-2024-8290, after a user fills in a form (\ref{fig:form}) and clicks the submit button, the browser generates an HTTP request with body payload (\ref{fig:http-request}) containing additional fields called sys-gen data: \textit{action}, \textit{controller}, \textit{status}, and \textit{wcfm\_ajax\_nonce}, is invisible in users' direct view. Furthermore, the field of \textit{wcfm\_customers\_manage\_form} also consists of some sys-gen data. 
    }
    \label{fig:sys-gen-example}
\end{figure}

For vulnerable object identification (C2), we first observe system-generated (sys-gen) data, which is restricted data that is invisible in users' direct view but visible in generated HTTP requests. Compared to user-supplied input that the web users can fully control, the sys-gen data is not controllable from the web UI. As explained in our analysis (Section \ref{root-cause-analysis}), BAC occurrences come from unexpected alterations in sys-gen data; thus, manipulating it can trigger BAC. In addition, OWASP released the top 25 vulnerable parameters \cite{OWASP_tp25param_2025}, and all of them are in the form of sys-gen data, making the sys-gen data an ideal mutation target. Web developers commonly place the sys-gen data in HTML files (e.g., in a hidden tag) or JavaScript files (e.g., in a function triggered on form submission). For example, Figure \ref{fig:sys-gen-example} demonstrates a filled-in form and an HTTP request generated by the web browser, in which there are additional (sys-gen) fields in the request that are invisible from the user's view.

\paragraph{\textbf{Data Collection}.}
\label{collection}
To collect sys-gen data, we analyze the intercepted requests stored in the corpus. Each key-value pair in the request URL or request body is classified as either the user-generated or sys-gen group by checking the value part. Since the fuzzer fills in all HTML forms with random values concatenated with certain pre-defined values, if a parameter value contains the pre-defined value, it is user-generated data; otherwise, it is sys-gen data.

\subsubsection{Reference Parameter Analysis by LLM}
\label{llm-analysis}
\begin{figure}[ht]
\centerline{\includegraphics[width=1\linewidth]{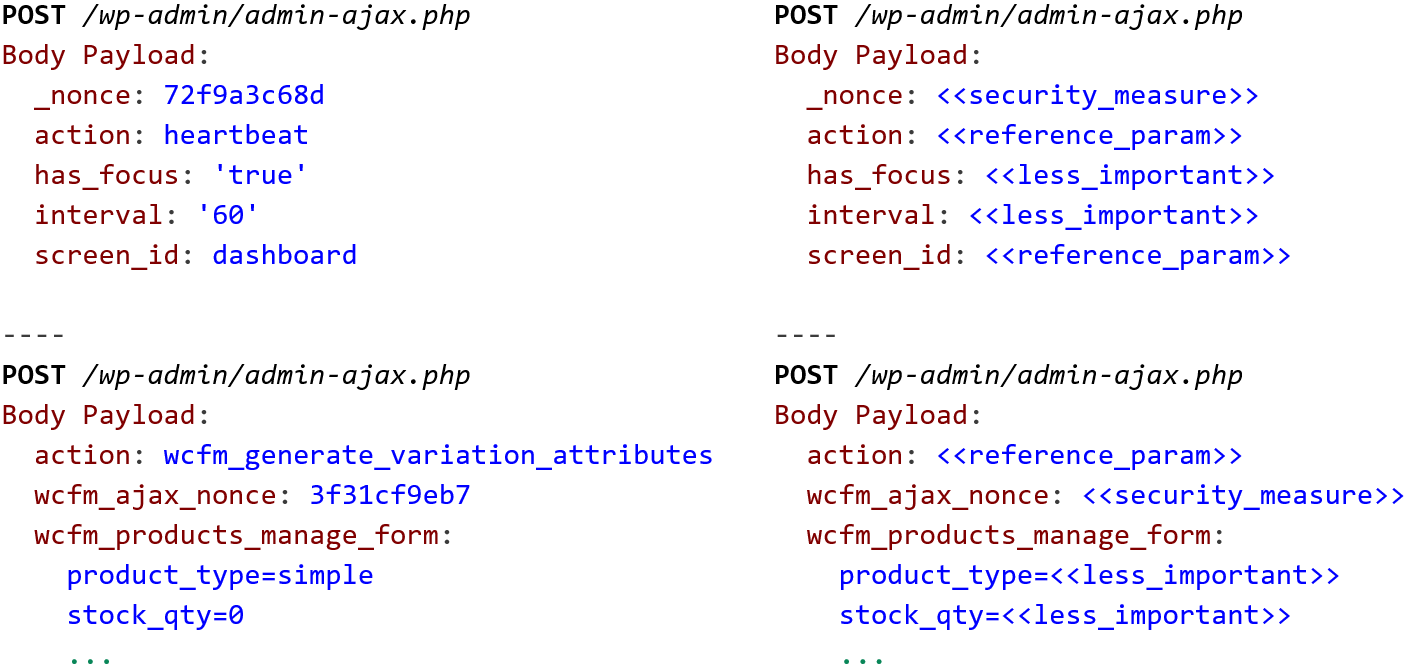}}
\caption{Two examples of HTTP requests after parameter analysis. The requests on the left are the original, and the right are the results.}
\Description{Two examples of HTTP requests after parameter analysis. The requests on the left are the original, and the right are the results.}
\label{fig:analysis-result}
\end{figure}

Since not all sys-gen data represents web functions or objects, we utilize LLM to further find out which sys-gen data refers to a function or an object. Therefore, after marking sys-gen parameters, the fuzzer queries LLM with a prompt described in Figure \ref{fig:LLM-prompt}. Getting replies with parameter names that may refer to functions or objects from LLM, the fuzzer marks them as \textbf{reference params} and special mutations are prepared for them. The other parameters are marked as less important (see Figure \ref{fig:analysis-result}), which means random mutation is always applied to them. According to their values or the name matching the predefined rules, parameters can be marked as security measures, which means the fuzzer will never select them for mutation.

\begin{figure}[!t]
\scriptsize
        \FramedBox{8.1cm}{0.45\textwidth}{
            \textbf{\# Overview} \\
You are a web fuzzer trying to reveal broken access control problems on web functions and objects.
I will give you one HTTP request consisting of the request method, URL, and body payload I have collected.
Based on this request, your goal is to find certain parameter names or body payload names that might refer to restricted objects.
Then, I will mutate the names you will give to examine if the mutated values for the names can access the restricted objects.
\\ \\
\textbf{\# Collected Requests} \\
I have crawled the web under test and caught many HTTP requests.
Now, I am giving you one request with a format of [method]<space>[full\_url]<space>[body\_payload].
\\ \\
\textbf{\# Your Task} \\
Based on that HTTP request, you have to point out which parameters or payload names that could potentially refer to restricted objects, excluding purely data-related fields or security parameters (e.g., nonces or security token).
Also, you have to point out which parameters or payload names that are reference IDs.
You might need to break down nested parameter values into their atomic components and only list their atomic parameters or payloads. Please note, only answer the question without explanation. Only use comma to separate the parameter or payload names.    
        }
    \caption{LLM Prompt for Parameter Analysis.}
    \Description{LLM Prompt for Parameter Analysis.}
    \label{fig:LLM-prompt}
\end{figure}

\subsubsection{Mutation}
\label{dictionary-mutation}
To reduce randomness (C4), the fuzzer should more often use values from the param corpus collection rather than random values for request mutation. 
Basically, reference values are grouped into two types: numeric and text, in which the former is any value starting with a number (can be followed by text) and the latter is otherwise. When the fuzzer takes an HTTP request and selects one of the reference parameters in the request for mutation, the value of the selected reference parameter is changed to one of the other reference values of the same type. It aims to reduce server rejection and increase mutation effectiveness since the reference values, especially the numeric ones, are usually limited and may refer to a certain object reference that has been seen before.


\textbf{Random Mutation.}
\label{random-mutation}
A random mutation that produces random values is still used and commonly applied to less-important parameters. When selecting an HTTP request for mutation, the fuzzer may only mutate selected reference parameters or also involve random mutation for selected less-important parameters.









\subsection{SQL Checking}
\label{SQL-checking}
For Oracle verification (C3), we propose SQL checking, which verifies whether WUTs execute the access attempt to protected references. While SQL queries alone may not be definitive proof of existence bugs, arbitrary values in data-manipulation queries indicate improper authorization checks. Illustrated in Figure~\ref{fig:req-query}, generally, a WUT generates DML (data manipulation language) to be sent to the DBMS when executing a request. To enable the real-time check, this study instruments the web interpreter to record all SQL queries the WUT generates. Then, the proposed fuzzer can check if the mutated values appear in the query. 

\subsubsection{Checking Rules}
Several rules are determined to automatically infer the occurrence of the BAC.


\paragraph{\textbf{Rule 1: Broken Function-level}}
Based on the BFLA definition, a web function is called vulnerable to access control if the function that only exists in certain roles is successfully executed by users operating under different roles. 
Therefore, when the fuzzer submits a request using a lower-role account and detects a DML query, it is \textbf{BFLA} if: 1) at least one value in the query is the same as the parameter value of the submitted request, and 2) the request does not exist in the user page. To ensure the second requirement is satisfied, the fuzzer opens the \textit{referer} link stated in the request header to check if the trigger source (e.g., button or link that may trigger the request) or the parameter values exist in the page.


\paragraph{\textbf{Rule 2: Broken Object-level}}
Based on the BOLA definition, an object is called vulnerable to access control if the object reference that only exists in certain roles is successfully accessed by users operating under different roles. The functions or URLs may exist in all roles, but they manipulate different objects. Therefore, when the fuzzer submits a request with altered reference params and detects a DML query, it is \textbf{BOLA} if: 1) at least one value in the \textit{WHERE} clause is the same as one of the mutated reference values in the corresponding request, and 2) the mutated reference value is not found in the corpus of the role that is used to submit the request. Due to the dynamic nature of the object, to ensure the second condition is satisfied, the fuzzer opens the \textit{referer} link stated in the request header, extracts the reference parameters from the opened page, and checks if the mutated values do not exist in the extracted parameter values.




\subsubsection{Multiple Checking for BAC Validity}
Using the SQL checking method, the fuzzer can detect a BFLA or BOLA; however, the fuzzer still needs to apply multiple checks to ensure that the values appearing in the SQL query are not by accident and indeed related to the mutated values in the submitted request. So, when the checking method marks a request as BAC according to the aforementioned rules, the fuzzer selects and mutates the request again. If the SQL checking detects the same BAC after the WUT executes the mutated request, then \textbf{the BAC is valid}. To increase the confidence level of BAC detection, the fuzzer selects and mutates the request again and again until the SQL checks do not raise BAC or the checking repetition reaches a certain number (e.g., 10 times repetition). 




\begin{figure}[!t]
    \begin{subfigure}[b]{0.5\textwidth}
        \FramedBox{1.3cm}{0.9\textwidth}{\textbf{\# Submitted HTTP Request}\\
            POST /lms/classes/Users.php?f=save 
            \\
            \\{id=\textcolor{red}{\textbf{1711440657}}\&firstname=ibdjweyxfoz\&middlename=wnmnozap ....}
        }
        \FramedBox{1.3cm}{0.9\textwidth}{\textbf{\# Generated SQL Query}\\
            \textit{UPDATE users set  firstname = 'ibdjweyxfoz'  ,  middlename = 'wnmnozap'  ,  lastname = 'swkgxicagowetfbp'  ,  username = 'tzqwyaloxx'  ,  password = 'c82a88d7c5f180798b0d118d23893a94'  where id = \textcolor{red}{\textbf{1711440657}}}
        }
    \end{subfigure}
    
    \caption{When a database-backed web app executes a request, it commonly generates an SQL query corresponding to the submitted data. For example, in CVE-2024-3139, the value of 1711440657 in the request payload appears in the SQL query. Therefore, the occurrence of certain values in an SQL query can be a sign that a certain request is executed, not rejected.}
    \Description{When a database-backed web app executes a request, it commonly generates an SQL query corresponding to the submitted data. For example, in CVE-2024-3139, the value of 1711440657 in the request payload appears in the SQL query. Therefore, the occurrence of certain values in an SQL query can be a sign that a certain request is executed, not rejected.}
    \label{fig:req-query}
\end{figure}

\section{\bacfuzz: Fuzzer Implementation}
\label{sec:implementation}
To adopt the proposed approaches, this study designs a new grey-box web fuzzer called \bacfuzz.

\begin{figure*}[ht]
\centerline{\includegraphics[width=0.8\linewidth]{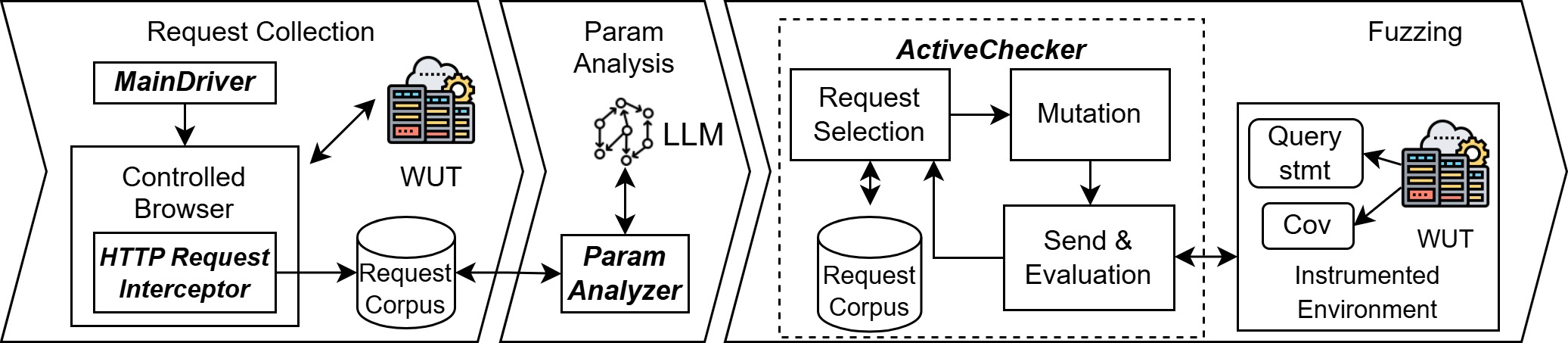}}
\caption{Proposed BACFuzz pipeline contains three main phases: request collection, parameter analysis, and fuzzing.}
\Description{Proposed BACFuzz pipeline contains three main phases: request collection, parameter analysis, and fuzzing.}
\label{fig:proposed-approach}
\end{figure*}

\subsection{Overview}
There are two main components of \bacfuzz: the \emph{main driver} and the \emph{active checker}. While the main driver navigates whole web pages in the WUT and stores submitted HTTP requests, the active checker processes the stored requests to exploit any BAC possibility in the WUT. 
Following the best practice of Witcher \cite{trickel_toss_2023}, we run both components sequentially, in which the main driver runs first.

In general, our proposed fuzzer (see Figure~\ref{fig:proposed-approach}) works in the following order: (1) collecting web functionalities and objects from each user role; (2) analysing reference parameters; and (3) altering the parameters and verifying the vulnerability occurrence. Initially, the fuzzer user sets the WUT URL and a list of registered accounts with various roles, including an anonymous account (see Algorithm~\ref{fuzzer-algorithm}). Iterating the roles, the fuzzer opens the URL and requires the user to open the login page and input the account credentials (line 4). After that, the driver works automatically to navigate whole web pages using the logged-in credentials (line 6). 
This process will be explained further in Section \ref{main-driver}. Then, the fuzzer determines the highest user role as the target. The active checker performs fuzzing using each user role (excluding the target), which will be explained further in Section \ref{active-checker}.



\begin{algorithm}[t]
\caption{BACFuzz}\label{fuzzer-algorithm}
\begin{algorithmic}[1]

\Require $baseUrl$
\Require $Roles$   \Comment{list of available roles}

\For{$r$ \textbf{in} $Roles$}
\State $state_r \gets userLogin(baseUrl, r)$
\State $d \gets $ \textbf{new} $MainDriver(state_r, r)$
\State \textbf{async} $d.crawl()$ \Comment{start the driver to navigate pages}
\EndFor

\State $target \gets getHighestRole(Roles) $ \Comment{e.g., Admin}
\State $Roles \gets excludeTarget(target,Roles)$
\State $resetWUT()$

\For{$r$ \textbf{in} $Roles$}
\State $state_r \gets userLogin(baseUrl, r)$
\State $c \gets $ \textbf{new} $ActiveChecker(state_r, r)$
\State \textbf{async} $c.fuzz()$ \Comment{start the active checker}
\EndFor
\end{algorithmic}
\end{algorithm}








\subsection{Main Driver}
\label{main-driver}
The main driver collects HTTP requests by navigating whole web pages using each user role. To enable the request collection, there are two main procedures in the main driver: \textit{Drive} and \textit{Intercept-Request}. The first procedure navigates the entire web page to find HTML pages manipulate web resources. The procedure looks for pages that have the \textit{form} tag and its derivatives (e.g., the \textit{input}, \textit{select}, and \textit{textarea} tags) because those allow users to input something to be sent to the server. On the other hand, the \textit{InterceptRequest} procedure aims to catch HTTP requests and store them in the request corpus. In addition, key-value parameters, in the request URL and body, are stored in the param corpus. We attached this procedure to the browser manipulation library, guaranteeing it is invoked whenever the browser sends any HTTP request.

\subsubsection{User Login}
To explore the whole web application functions, the main driver requires the fuzzer user to log in using some accounts with different roles. For example, if the user intends to test WordPress, which consists of five roles: administrator, editor, author, contributor, and subscriber, the driver will prompt the user to log in five times. This login process can be done automatically, which requires the user to configure login scripts to align with the WUT page. After storing all cookies and the user URLs (dashboard pages) \cite{olsson_spider-scents_2024}, the fuzzer instantiates some driver instances to navigate the main page using the cookies.

\subsubsection{Crawling Strategies}
According to the work of Stafeev and Pellegrino \cite{stafeev_sok_2024}, which reviews many crawling algorithms, there are two crucial problems that need to be specified when designing a web crawler: navigation and page similarity.

\paragraph{Navigation strategy.}
Similar to the work of Khodayari et al. \cite{khodayari_great_2024} which crawls around 1M web pages, our main driver visits web pages with a depth-first strategy, which means visiting the most recently discovered link. Given the homepage of the logged-in user, the driver identifies web links by looking for anchor tags and saves them in the \textit{links} variable. 


\paragraph{Page similarity.}
The proposed driver uses URL components as the objects for page similarity, which means only visiting pages with different URL components. When finding a web link from an anchor tag, the driver takes its visible text and URL components (i.e., base URL, URL path and query string containing sys-gen data) and compares this information with previously saved links. If it is non-identical, the found link is stored and will be visited.

\subsection{Active Checker}
\label{active-checker}
The main idea of the active checker is to alter the captured HTTP requests to reveal BAC. In general, the active checker chooses one HTTP request from the seed (i.e., request corpus), performs one of the mutation strategies, sends the mutated requests to the WUT using a certain user session, and checks the feedback. 

\subsubsection{Request Selection}
\label{attack-surface-selection}
The active checker prioritises HTTP requests with more reference parameters to be selected because the endpoints of those requests have more attack surfaces that can be exploited. To implement that idea, the request selection process involves a random function with the number of reference params as the weight. 

\subsubsection{Mutation}
After obtaining a request for mutation, the active checker chooses what kind of testing (either function- or object-level) to conduct based on the request mark (explained in Section~\ref{request-interception}).



\paragraph{Function-level Testing.}
BFLA happens when the WUT execute HTTP requests that are unavailable for certain user roles. 
For example, the request of \textit{add\_item} that only exists on the administrator page will be tested using a non-admin account. So, when taking an HTTP request having a different label from the checker name (explained in Section \ref{collection}), the active checker will perform such testing because the request is supposed to be rejected by the WUT due to forbidden access. To be more effective, the active checker adjusts the chosen request with unique data from the selected accounts, such as \textit{nonce} data that is unique per account in WordPress.



\paragraph{Object-level Testing.}
BOLA mostly happens when a certain role is allowed to execute certain functions but the WUT does not validate the object the user tries to modify. Therefore, to test the object level, the checker takes one request with the same label as the checker's logged-in role and mutates the selected parameters by using another object reference value (Section \ref{dictionary-mutation}).

\subsubsection{Feedback Evaluation}
\label{feedback-monitoring}
The feedback the active checker receives from the WUT (i.e., HTTP response code, response messages, code coverage, and SQL queries) is evaluated to decide whether the sent request has triggered BAC. As explained in Section~\ref{SQL-checking}, the active checker compares parameter data inserted in the submitted requests with the SQL query the WUT produces. If the data exists, the request and the query are reported to the user and added to the attack surface collection. In addition, when requests fail to trigger BAC yet bring new code coverages or 500-response codes, the active checker also puts these requests into the attack surface collection to be explored further. As shown in a previous study \cite{van_rooij_webfuzz_2021}, these interesting requests can guide the fuzzer to reach deeper statements in WUTs. We count the 500-response code as well because some studies \cite{corradini_automated_2023}\cite{du_vulnerability-oriented_2024} stated that this code is useful to show web defects and can lead to more vulnerabilities.





\subsection{Instrumentation}
\label{instrumentation}
Instrumentation is critical in grey-box fuzzer setup because it produces live and lightweight internal information \cite{manes_art_2021}. This study follows the work of Neef et al \cite{neef_what_2024} that instruments web applications by using the \emph{function hooking}, a feature provided by the UOPZ library \cite{noauthor_phpuopz_nodate}, to collect SQL queries sent by WUT. Using this technique enables the instrumentation scripts to manipulate original PHP functions related to SQL calls, such as \textit{mysqli\_query}, \textit{mysqli\_stmt\_prepare}, \textit{mysqli\_prepare}, and \textit{PDOStatement} with additional codes acting as query catching.
In addition to the query collection, we use the PCOV  library \cite{watkins_krakjoepcov_2025} for coverage accounting. Since the library has a better performance overhead than Xdebug, it helps generate accurate line coverage reports quickly \cite{PCOVorXdebug_2024}. 
Since each request is supplied with a unique identifier (named \textit{X-FUZZER-COVID}) attached in the header, both query and coverage information are written in JSON files named with each corresponding request identifier. 

\begin{table*}[htbp]
\caption{Evaluation results sorted by the number of WUT LoC. It shows that BACFuzz can detect 16 out of 17 known cases and reveal 26 new TP BAC. The only failed detection occurred in CVE-2023-43663 due to discrepant submitted values.}
\vspace{-0.3cm}
\small
\begin{center}
\begin{tabular}{c|p{4cm}|c|c|c|c|c|c|c|c||c|c|c}
\hline
\textbf{\makecell{App Name\\/ CVE No}} & \textbf{\makecell{Available Roles\\(apart from Admin}} & \multicolumn{3}{c|}{\textbf{\makecell{Number of}}} & \textbf{\makecell{Req\\Col.}} & \textbf{\makecell{Known\\BAC}} & \textbf{\makecell{Detect\\Time}} & \multicolumn{2}{c||}{\textbf{\makecell{New BAC\\Detected}}} & \textbf{\makecell{Avg.\\Resp.}} & \textbf{\makecell{Instr.\\Over}} & \textbf{\makecell{Non-\\Rej.}}\\
\cline{3-5} \cline{9-10}
&  \makecell{\textbf{and Anonymous)}} & Files & LoC & Req & \textbf{Time} & & & TP & FP & \textbf{Time} & \textbf{head} & \textbf{Req.}\\
\hline
DVWA & User & 0.2K & 7.5K & 71 & 0:04:15 & \makecell{BFLA} & 0:00:02 & 0 & 0 & 0.025 & 0.001 & 99\% \\
\hline
XVWA & User & 0.7K & 17K & 44 & 0:26:33 & \makecell{BFLA} & 0:00:15 & 0 & 0 & 0.026 & 0.003 & 100\%\\
\hline
\hline
CVE-2025-0843 & User & 17 & 0.5K & 13 & 0:01:19 & BFLA & 0:00:02 & 1 & 0 & 0.021 & 0.003 & 100\% \\
\hline
CVE-2025-3536 & \multirow{2}{*}{Employee} & \multirow{2}{*}{40} & \multirow{2}{*}{1.5K} & \multirow{2}{*}{63} & \multirow{2}{*}{0:04:42} & BFLA & 0:00:02 & \multirow{2}{*}{1} & \multirow{2}{*}{0} & \multirow{2}{*}{0.023} & \multirow{2}{*}{0.003} & \multirow{2}{*}{100\%}\\
\cline{1-1}\cline{7-8}
CVE-2025-3537 & & & & & & BFLA & 0:01:25 & & & & \\
\hline
CVE-2024-55231 & \multirow{2}{*}{User} & \multirow{2}{*}{0.3K} & \multirow{2}{*}{2.4K} & \multirow{2}{*}{38}& \multirow{2}{*}{0:02:33} & BOLA & 0:00:06 & \multirow{2}{*}{0} & \multirow{2}{*}{0} & \multirow{2}{*}{0.022} & \multirow{2}{*}{0.004} & \multirow{2}{*}{99\%}\\
\cline{1-1}\cline{7-8}
CVE-2024-55232 & & & & & & BOLA & 0:00:28 & & & & \\
\hline
CVE-2024-40480 & User & 0.1K & 4K & 78 & 0:06:24 & BFLA & 0:00:06 & 2 & 0 & 0.024 & 0.007 & 99\%\\
\hline
CVE-2025-0802 & LeaveMan., Salary, BorrowMan. & 0.6K & 23K & 124 & 0:15:37 & BFLA & 0:01:54 & 3 & 0 & 0.022 & 0.002 & 100\%\\
\hline 
CVE-2023-46449 & Staff & 1.1K & 145K & 67 & 0:12:14 & BOLA & 0:01:36 & 8 & 0 & 0.021 & 0.003 & 100\%\\
\hline
CVE-2024-3139 & Staff & 2.5K & 149K & 28 & 0:03:12 & BFLA & 0:00:08 & 2 & 0 & 0.018 & 0.001 & 99\%\\
\hline
CVE-2024-9082 & Staff & 2.6K & 152K & 71 & 0:18:13 & BFLA & 0:00:52 & 2 & 0 & 0.024 & 0.002 & 100\% \\
\hline
CVE-2024-7658 & Manager, Uploader, Client & 12.3K & 159K & 642 & 3:05:02 & BFLA & 0:0:28 & 0 & 0 & 0.091 & 0.039 & 61\%\\
\hline
CVE-2024-7437 & \multirow{2}{*}{User} & \multirow{2}{*}{0.8K} & \multirow{2}{*}{188K} & \multirow{2}{*}{1458} & \multirow{2}{*}{2:49:37} & \makecell{BOLA} & 0:06:46 & \multirow{2}{*}{1} & \multirow{2}{*}{5} & \multirow{2}{*}{0.092} & \multirow{2}{*}{0.041} & \multirow{2}{*}{73\%} \\
\cline{1-1}\cline{7-8}
CVE-2024-7438 & & & & & & \makecell{BOLA} & 0:04:21 & & & & & \\
\hline
CVE-2024-8290  & ShopMan., Author, Subscriber, Cust. & 9.3K & 929K & 548 & 3:10:04 & \makecell{BOLA} & 0:03:06 & 4 & 2 & 1.423 & 0.864 & 54\%\\
\hline
CVE-2023-43663 & Logistician, Translator, Sales, Cust. & 29.3K & 2.4M & 1897 & 4:09:38 & BFLA & X & 1 & 3 & 0.298 & 0.162 & 99\%\\

\hline
\hline
\makecell{OpenCart} & Cataloger, Marketing, Customer & 9.8K & 434K & 288 & 0:30:34 & - & - & 0 & 1 & 0.105 & 0.025 & 55\%\\
\hline
\makecell{ZenCart} & Order Processor, Customer & 9.7K & 585K & 878 & 2:37:27 & - & - & 1 & 0 & 0.476 & 0.131 & 91\%\\
\hline
\makecell{phpBB} & Member & 7.8K & 726K & 543 & 1:10:39 & - & - & 0 & 0 & 0.101 & 0.052 & 97\%\\
\hline
\end{tabular}
\label{tab-benchmark-evaluation}
\end{center}
\caption*{
  \parbox{\textwidth}{
    \centering\small
    Req Col Time = Time to collect all unique HTTP requests; 
    Avg Resp Time = Average time of WUT, which has been instrumented, to reply;\\
    Instr Overhead = Time overhead caused by instrumentation; 
    Non-Rej Req = Proportion of requests that are not rejected by WUT.
  }
}
\vspace{-0.5cm}
\end{table*}

\subsection{Counting Unique Results}
After running a certain of time, \bacfuzz~ reports the final results to the user. To count unique BAC cases, \bacfuzz~ compares the submitted request URL and method with the detected SQL query. For example, 
CVE-2024-7437 and CVE-2024-7438 use the same URL and method to trigger the vulnerability; however, the SQL queries produced are different (i.e., \textit{DELETE FROM smf\_user\_alerts} vs \textit{UPDATE smf\_user\_alerts}), making them two different cases.

\subsection{Implementation}
We implemented the proposed fuzzer in Python because it has libraries that fit our needs, such as Playwright \cite{playwright_2025} for the browser controller and request interceptor, and BeautifulSoup \cite{beautifulsoup4} for HTML processing. Then, we implemented the instrumentation, including the SQL checking, in PHP scripts with the UOPZ and PCOV libraries. To ease the evaluation process, we deploy the WUTs and evaluation scripts in Docker because handling complex web server components is straightforward with Docker. In addition, we use Docker to make it easy for other researchers to duplicate our study. 

\section{Empirical Evaluation}
In order to evaluate the effectiveness of \bacfuzz~ comprehensively, we run it against test-bed and real web applications with known and unknown access control vulnerabilities. At the end of the experiments, we aim to answer the following research questions.


\begin{itemize}[leftmargin=*]
    \item RQ1. Can \bacfuzz~report the known vulnerabilities?
    \item RQ2. Can \bacfuzz~uncover new and valid vulnerabilities?
    \item RQ3. How much overhead does \bacfuzz~introduce?
    \item RQ4. How effective is \bacfuzz~in generating tests?
\end{itemize}


\subsection{WUT Collection}
As WUTs with known vulnerabilities, we use web applications affected by 15 CVEs summarized in Section \ref{root-cause-analysis}. 
We also include benchmark applications employed by recent studies \cite{li_fuzzcache_2024}\cite{neef_what_2024}\cite{guler_atropos_2024} that are PHP-based: \textit{Damn Vulnerable Web Application} (DVWA) and \textit{Xtreme Vulnerable Web Application} (XVWA), since they provide BAC vulnerabilities for training purposes.



Recently, a static analysis work \cite{huang_detecting_2024} revealed BOLA vulnerabilities in 25 web applications. Even though this work successfully detects many vulnerabilities in most tested applications, it finds nothing in four applications: Scarf, PhpBB, Opencart, and Zencart. Except for Scarf, which is no longer updated, we also evaluate our work on those three remaining applications to demonstrate that our fuzzer can reveal vulnerabilities that the static analysis work cannot detect. 

\subsection{Experimental Setup}
\label{sub:exsetup}
For each WUT, we run the \bacfuzz~main driver for a maximum of 24 hours to collect all HTTP requests and continue to run the active checker for 24 hours to reveal BAC from the requests. We run the experiments in a virtual computer with 8 CPUs and 32 GB RAM and use the \textit{DeepSeek-V3} model provided by \textit{DeepSeek} \cite{noauthor_deepseek_nodate} for LLM service because it has good performance in the code generation domain. Since each fuzzing campaign works randomly, making the result not the same in each experiment, we run the fuzzer three times and report the average results.

In the experiments, we do not compare our work with other popular security tools, such as OWASP ZAP and Burp Suite, due to the significant difference between functionality and scope. Unlike our approach, they are not designed for fully automated, object-level access control testing, making them not applicable for a direct comparison. In addition to those popular tools, there are several scientific works addressing BAC, such as \cite{rennhard_automating_2022}\cite{kushnir_automated_2021}\cite{sun_static_2021} (explained more in Section \ref{related-work}); however, the source codes of these works are not available, preventing us from comparison.
There is also a static analysis work \cite{huang_detecting_2024} to reveal BOLA vulnerabilities, but the authors confirmed that it is challenging to manually create \textit{dal\_specification.json}, the required file for testing new WUTs.

\subsection{Experiment Results}
Answering RQ1, \bacfuzz~successfully reported 16 out of 17 known vulnerabilities (see Table \ref{tab-benchmark-evaluation}). The only failed detection happens in the CVE-2023-43663 case due to discrepant submitted values. In that case, which affects Prestashop, when the disable module function is called, the WUT converts the module ID to a corresponding primary-key ID in the DBMS and then sends the converted ID through a SQL query. For example, \bacfuzz~sends \textit{statsbestcustomers} in the vulnerable parameter, but the WUT sends the query of "\textit{UPDATE ps\_module SET active = 0 WHERE id\_module = 64}" to DBMS, making the fuzzer unable to detect the matched value.

Answering RQ2, \bacfuzz~ successfully reported new vulnerabilities with low false positive rates. \bacfuzz~also reported one new BAC from applications that the static analysis \cite{huang_detecting_2024} could not detect. We reported these vulnerabilities to the developers of the respective applications and are waiting for the developers' responses to confirm our findings.
If we receive confirmation before the final version deadline, we will include their responses in the revision. Otherwise, we will clarify the disclosure timeline and our communication efforts in the paper.



To answer RQ3, we compare the time WUTs need to reply to each request between those using and not using instrumentation. We deactivate the instrumentation by not putting the \textit{X-FUZZER-COVID} header in the request (as explained in Section~\ref{instrumentation}). Table \ref{tab-benchmark-evaluation} shows that the use of instrumentation for coverage and query collection results in less overhead.

To answer RQ4, we compare the number of not-rejected requests (replied with either 200 or 500 response codes) because those requests are interesting, as explained in Section ~\ref{feedback-monitoring}. The results in Table \ref{tab-benchmark-evaluation} show that the proposed mutation strategies can produce a large proportion of non-rejected requests, which is good for exploiting various logic functions in WUT.

\subsection{Discussion: False Positive Result}
Although SQL query can serve as an oracle for detecting BAC, its application may also result in false positive (FP) results. 
Based on our observation of evaluation results, FP happens because of the coincidence of the same value and the nature of dynamic objects.

First, the matched value between the mutated parameter and the captured SQL query is syntactically valid yet semantically incorrect. For example, in OpenCart, \bacfuzz~collects a query of "\textit{DELETE FROM oc\_cart WHERE (api\_id > 0 OR customer\_id = 1)}" after sending a request with \textit{route=0} as the mutated parameter value. Since this condition is marked as BAC due to the presence of 0 value, multiple checks are performed by sending a new mutated request with \textit{route=1}. Once again, the condition is marked as BAC because the value 1 exists in the collected query. However, we can obviously see that the submitted request and the obtained query are not correlated.

Second, dynamic object behaviour causes FP because certain objects are missing during crawling, but then exist during fuzzing. For example, in SMF, a user had only one readable topic on his page initially, making the MainDriver store the \textit{id\_topic} 1 as the available object for the user. However, during the fuzzing campaign, the ActiveChecker might send requests that lead to another topic creation (e.g., \textit{id\_topic} 2) for the user. Therefore, because \textit{id\_topic} 2 is still considered inaccessible but \bacfuzz~ successfully modifies it (because it is indeed accessible for the user), \bacfuzz~ raises a BOLA flag that is FP.

\section{Threats to Validity}
While this study demonstrates the effectiveness of \bacfuzz~in revealing BAC vulnerabilities, some threats to validity must be considered when interpreting the results.

\paragraph{\textbf{Internal Validity}}
The internal threat arises from how large WUTs, such as SMF, WordPress, and PrestaShop, handle logging, as they store extensive runtime events in dedicated log tables. These log tables commonly record errors and unauthorized access attempts, making SQL queries mostly match the submitted requests. Therefore, it is suggested to exclude this kind of table that always stores all submitted requests. To make the fuzzer only report valid vulnerabilities, the fuzzer user should put these ignored tables in the \bacfuzz~ configuration. These ignored tables are crucial to maintaining the report's precision. In our experiment, we exclude tables of \textit{smf\_log} (CVE-2024-7437 and CVE-2024-7438) and \textit{ps\_connections} (CVE-2023-43663) from the fuzzer observation. While this filtering step improves the precision of our results, it also introduces a slight dependency on user knowledge of the WUT.

\paragraph{\textbf{Construct Validity}}
The construct validity threat comes from our BAC dataset. While additional related CVEs may exist beyond our collection, we believe the selected cases are sufficiently diverse to support the design of our fuzzer. The dataset covers a wide range of BAC patterns and application behaviours, making it a representative basis for guiding and evaluating our design. Nonetheless, our collection methodology may introduce bias toward well-documented or easily reproducible vulnerabilities, which may not capture all real-world scenarios.


\section{Related Work}
\label{related-work}
We identified some prior works related to \bacfuzz, especially in the web fuzzing domain and the access control testing domain.

First, there are some state-of-the-art web fuzzers, such as \textit{EvoMaster} \cite{arcuri_evomaster_2018}, \textit{Restler} \cite{atlidakis_restler_2019}, \textit{RestTestGen} \cite{viglianisi_resttestgen_2020}, \textit{bBOXRT} \cite{laranjeiro_black_2021}, and \textit{RESTest} \cite{martin-lopez_restest_2021}, which work effectively in triggering web crashes. However, they cannot reveal BAC problems because the characteristics between crash and BAC are very different. There are also vulnerability-driven fuzzers which focus more on non-crash vulnerabilities, like \textit{Witcher} \cite{trickel_toss_2023}, \textit{EDEFuzz} \cite{pan_edefuzz_2024}, \textit{ResolverFuzz} \cite{zhang_resolverfuzz_2024}, and \textit{Atropos} \cite{guler_atropos_2024}; however, they are not revealing BAC.

Second, there are some recent works on access control testing. Rennhard et al. \cite{rennhard_automating_2022} and Kushnir et al. \cite{kushnir_automated_2021} developed practical solutions to automatically detect BAC on web applications. Since their works are limited by manipulating GET requests only, performed in a black-box setup, and comparing web responses for vulnerability verification, these are different from our work. Our proposed fuzzer manipulates more attack vectors, works in a grey-box setting with more advanced techniques, and uses SQL queries to verify the vulnerability. The work of Sun et al. \cite{sun_static_2021} used static analysis for vulnerability detection; however, it only works well for traditional web applications that are not AJAX-heavy. Our work uses a dynamic testing setup that catches HTTP requests, rather than observing HTML links, thus it can handle modern web applications. 
As mentioned in Section~\ref{sub:exsetup}, existing work based on static analysis \cite{huang_detecting_2024} requires manual creation of test files, making it less practical than our work.
Lastly, the work of Kim et al. \cite{kim_finding_2020} investigates parameter tampering on the web; however, they highly rely on human intervention to decide whether tampering is successful in altering the business process. Our fuzzer uses automatic verification through query checking, making much less human involvement.

In the domain of vulnerability scanners, there are two widely used tools: OWASP ZAP \cite{noauthor_owaspzap_nodate} and Burp Suite \cite{noauthor_burp_nodate}, which operate in a black-box setting to detect various security issues, including BAC. However, their access control testing is not fully automated. Users are required to manually configure an access control list (ACL) defining which functions should be allowed/denied for specific users. This setup process is both time-consuming and challenging. Furthermore, those tools do not support object-level access control testing and rely solely on web responses to verify ACL violations. As discussed in Section~\ref{oracle-verification}, this verification method is often insufficient for accurately identifying authorization flaws.
To address these limitations, our work introduces automated approaches that eliminate the need for predefined ACLs and accurately verify BAC.

\section{Conclusion and Future Work}

Broken Access Control (BAC) vulnerabilities remain pervasive in web applications, yet pose unique challenges for automated detection due to the lack of reliable oracles and the difficulty of generating semantically valid attack inputs. In this paper, we presented \bacfuzz, a grey-box fuzzing framework specifically designed to uncover BAC vulnerabilities—including BOLA and BFLA—by combining hierarchical role analysis, reference mutation, and SQL-based oracle checking. Empirical evaluation across 20 real-world PHP applications demonstrates that \bacfuzz~effectively detects 16 of 17 known issues and uncovers 26 previously unknown vulnerabilities, all with low false positive rates. By releasing the source code and curated evaluation dataset, we aim to foster further research on BAC vulnerabilities and support the development of more secure web applications.

As acknowledged in Section~\ref{scope-of-work}, there are some types of BAC that fall outside the scope of this work and represent directions for future exploration.
First, we identified context-dependent BAC, which refers to vulnerabilities that only manifest after a user performs specific actions, causing a WUT to enter a certain state. These cases require preconditions (e.g., resource creation or feature activation) before unauthorized access becomes observable.
Second, we identified passive or view-type BAC, which allows unauthorized users to gain access to sensitive information without modifying any data in the DBMS. Since \bacfuzz~relies on data manipulation (DML) queries to infer unexpected actions, it is difficult to verify the violation of restricted information displayed on user pages that only involve \textit{SELECT} statements. As a result, both context-dependent and passive BAC remain open challenges for future work.

\bibliographystyle{ACM-Reference-Format}


\end{document}